\documentclass[aps,twocolumn,nofootinbib]{revtex4}
\usepackage{graphicx}
\usepackage{adjustbox}
\usepackage{dcolumn}
\usepackage{bm}
\usepackage{amsmath}
\usepackage{multirow}
\usepackage{amsmath}
\usepackage[letterpaper, hmargin = {2cm}, vmargin = {2cm}]{geometry}
\usepackage{tablefootnote}
\usepackage[ulem=normalem]{changes}
\usepackage{mathrsfs}

\begin{document}

\title[Weak binding effects on   $^{40}$Mg -  Macchiavelli, Fallon, Crawford]{Weak binding effects on the structure of  $^{40}$Mg}

\author{A.~O.~Macchiavelli,  H.~L.~Crawford, P.~Fallon, and R.~M.~Clark}
\affiliation{Nuclear Science Division, Lawrence Berkeley National Laboratory, Berkeley, CA 94720, USA}

\author{A.~Poves} 
\affiliation{Departamento de F\'isica Te\'orica  and IFT-UAM/CSIC, \mbox{Universidad
 Aut\'onoma de Madrid, 28049 Madrid, Spain}}

\date{\today}

\begin{abstract}
While the phenomenon of one- and two-neutron ground-state halo nuclei is well established, the effects of weak binding on nuclear excitation properties remain largely unexplored. Motivated by this question and by recent data in $^{40}$Mg we investigate the coupling of weakly bound (halo) valence neutrons to a core using the known properties of $^{40}$Mg to explore and illustrate possible particle-core coupling schemes and their impact on the low-lying excitation spectrum.
\end{abstract}

\maketitle

\section{Introduction}

The properties of nuclei close to the neutron drip-line can differ from those near to stability due to the presence of weakly bound valence neutrons. One of the most striking structural changes to emerge is the formation of a neutron halo, where the outermost one or two weakly bound valence neutron states form a spatially extended surface comprised almost entirely of neutrons. As well as being near threshold the valence neutrons must also occupy low-$\ell$  states $(s, p)$ to minimize the effects of an angular momentum barrier and allow their wavefunction to extend significantly beyond the core.  

Focusing on two-neutron halos, first evidence came from a measurement of the interaction cross-section of $^{11}$Li~\cite{Tanihata85} reported in 1985. In their seminal work~\cite{Hansen87} Hansen and Jonson showed that the low binding in $^{11}$Li  leads to {\sl neutronization} of the nuclear surface as originally suggested by Migdal~\cite{Migdal73}, where the nucleus is surrounded by a neutron halo extending to several times the nuclear (core) radius.  It is interesting to note that this state may be interpreted as a bound dineutron coupled to the nuclear core~\cite{Migdal73}.  The extra attraction  between the two neutrons in the presence of a Fermi sea gives rise to a bound $nn$ structure despite the fact that the free di-neutron is not bound~\cite{Barranco01}.


While the phenomenon of a one- and two-neutron ground-state halo is well established (there are over 10 known cases that have been experimentally confirmed, ranging from $^{6}$He to $^{37}$Mg) the effects of weak binding on nuclear structure properties remain largely unexplored. Motivated by this question and by recent data in $^{40}$Mg~\cite{Crawford19} we investigate the coupling of weakly bound (halo) two valence neutrons to a core using the known properties of $^{40}$Mg to explore and illustrate possible particle-core coupling schemes and their impact on the low lying excitation spectrum. $^{40}$Mg lies close to the neutron drip line and is a potential candidate for a halo nucleus with two neutrons occupying the  $p_{3/2}$ orbit~\cite{Caurier14,Hamamoto07,Nakada18}, and where the observed spectrum of $\gamma$-ray transitions deviates from that seen in the neighboring more bound $^{36,38}$Mg isotopes. 

Of particular interest are the effects of the continuum on nuclear rotational motion.  These have been studied in Ref.~\cite{Fossez16A,Fossez16B} in the framework of the particle-plus-core problem using a non-adiabatic coupled-channel formalism.  The subtle interplay between deformation, shell structure, and continuum coupling can result in a variety of excitations in an unbound nucleus just outside the neutron drip line, as predicted for the low-energy structure of $^{39}$Mg. 

In this work we take a more qualitative view, yet one that captures the main physical ingredients, to study the low-lying excitations that may emerge in a halo nucleus.  We begin in Section II by discussing a ``universal'' plot that relates the $2n$ separation energy ($S_{2n}$) to the volume overlap between the core and halo as a way to characterize and identify possible halo nuclei, and then use the Weak Coupling~\cite{Arima71} and Particle-Rotor Models~\cite{ShapesAndShells} in a phenomenological approach in Section III, expanding on the scenarios first presented in Ref.~\cite{Crawford19}. 


\section{Core-Valence Overlap}

As discussed in Ref.~\cite{Hansen87} and subsequent works~\cite{Hansen95,Riisager92,Riisager13,Jensen03} (see also the recent monograph~\cite{Bhasin21}) the extended matter radius exhibited by a two-neutron halo nucleus can be expressed in terms of the separation energy of the weakly bound neutrons ($S_{2n}$) through the tunneling parameter $\mathscr{X}= r_{c} \sqrt{2\mu S_{2n}}/\hbar $ derived from the exponential nature of the asymptotic wavefunction. We thus consider a plot of $(<r_c^2>/<r_m^2>)^{1/2}$ vs. $\mathscr{X}$  as a way to capture the universal features of the 2$n$ halo systems, where $\mu$ is the reduced mass of the core + 2n system, and $r_c$, $r_m$ are the core and matter radii respectively.  Their root-mean-square (RMS) ratio represents the volume overlap between the valence nucleons and core.

\begin{figure*}[htbp]
\centering
  \includegraphics[width=\textwidth]{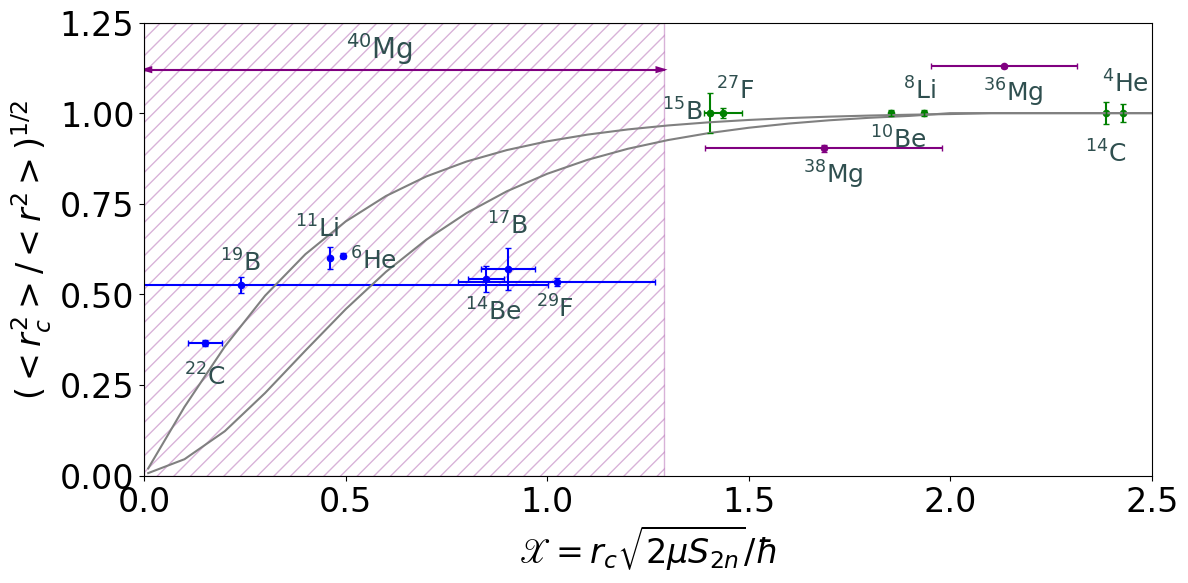}
  \caption{Universal plot for $2n$ halos showing the dependence of $(<r_c^2>/<r^2>)^{1/2}$, a measure of the volume overlap, with the tunneling parameter $\mathscr{X}$.  The black curves show the expected behavior for $\ell=0,1$.  The blue data points correspond to known $2n$ halo nuclei, while the green data points were used for fitting the value of $r_{c}$.  The purple data points correspond to the Mg isotopes.}
  \label{fig:2nHalos}
\end{figure*}

The experimental data on known $2n$ halo nuclei were obtained by comparing matter radii from Ref.~\cite{Ozawa01} to a local $A^{1/3}$ fit 
of $r_c$ in an isotopic chain for $\mathscr{X} > 1$.
The results are shown in Fig.~\ref{fig:2nHalos}, where the two black curves are the expectation for $\ell=0,1$.
The range of binding energy expected for $^{40}$Mg is indicated by the shaded region. The mass table evaluation~\cite{AME2021B,AME2021C} gives an expected $S_{2n}$ of $667 \pm 706$ keV. 
Within this range of binding energy we may expect a reduced core-valence overlap, suggesting that $^{40}$Mg is a possible $2n$ halo candidate where, as we shall discuss, a (almost) spherical $p^2$ two-neutron halo may develop on top of a deformed $^{38}$Mg core~\cite{Nakada18}.
An inspection of the figure clearly suggests that in the future both mass and interaction cross-section measurements of $^{40}$Mg will be crucial to firmly establish its (potential) halo nature.

\section{Particle-Core Coupling}

The spatial separation between the two halo neutrons and the core suggests that the motion (excitation modes) of the system as a whole is reasonably described by that of the individual subsystems and the strength and nature of the coupling between them. Here we discuss the strength of the interaction between the core and the 2n system in terms of particle-surface (phonon) coupling as a metric to establish whether the halo can be considered ``weakly'' or ``strongly'' coupled.

As discussed in detail by Bohr and Mottelson~\cite{Bohr53,BMVolume2}, the strength of the particle-vibration (surface) coupling for quadrupole modes, specifically relevant to the present discussion,  is given by:
\begin{equation}
\label{eq:formFactor}
f_{\lambda=2}= \left(\frac{5}{16\pi}\right)^{1/2}\left(\frac{\hbar\omega_2}{2C_2}\right)^{1/2} \frac{\langle k_2 \rangle}{\hbar\omega_2}
\end{equation}
\noindent
where $C_{2}$ is the quadrupole restoring force parameter, $\hbar\omega_{2}$ is the vibration frequency, and $k_{2}(r) = R~\partial V/\partial R$ the single-particle form factor, where we take $R$ as the core radius $r_{c}$. 
 In turn we have:
\begin{equation}
\label{eq:formFactor}
B(E2,0^+ \rightarrow 2^+)=5\big(\frac{3}{4\pi} ZeR^2\big)^2 \frac{\hbar\omega_2}{2C_2} \nonumber
\end{equation}

\begin{figure}[htbp]
\centering
  \includegraphics[width=\columnwidth]{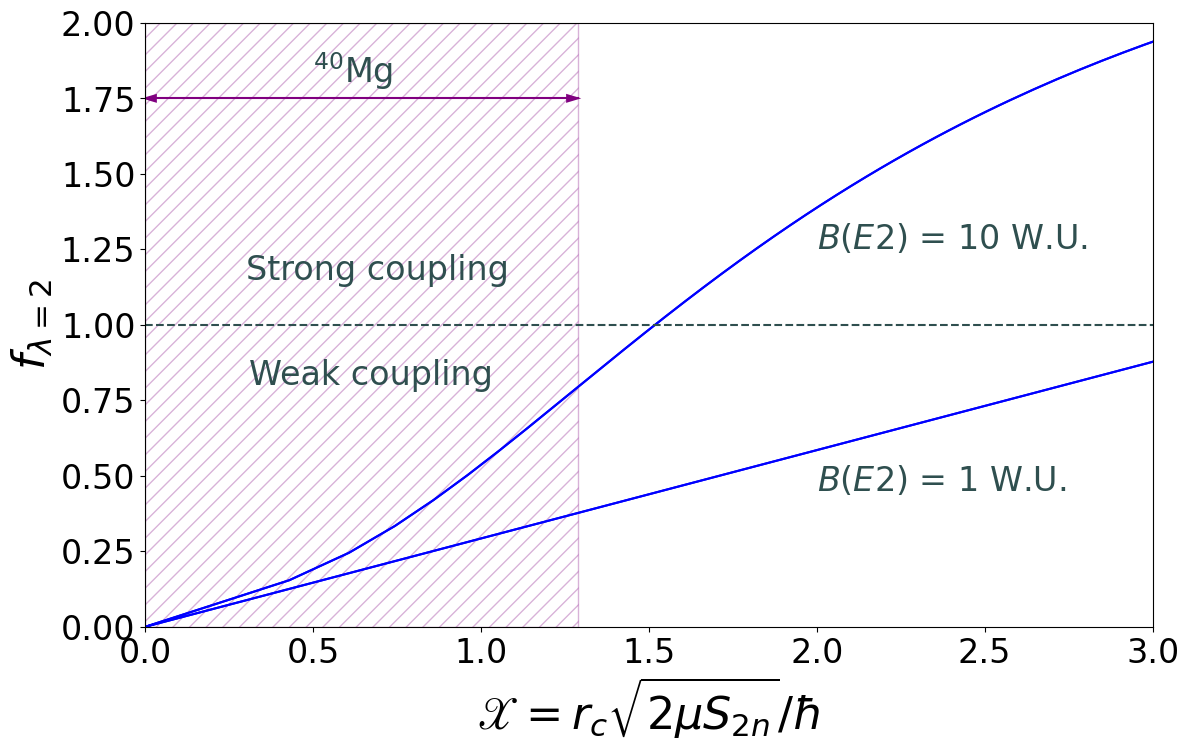}
  \caption{Bohr-Mottelson particle-surface quadrupole coupling form factor, $f_{\lambda=2}$, as a function of $\mathscr{X}$. Two cases are shown, corresponding to soft- and hard- cores as represented by the given $B(E2)$ strength in single-particle units.}   
  \label{fig:formFactor}
\end{figure}

Values of $f_{\lambda=2}$ greater than unity are associated with strong coupling while values of $f_{\lambda=2}$ less than unity are associated with weak coupling. The softness of the surface quadrupole vibration/phonon mode is related to the $C_2$ coefficient, or the $B(E2)$, and the frequency $\hbar\omega_2$.\footnote{In this context, for example,  the $^{9}$Li core in $^{11}$Li can be considered to be ``soft'', whereas the $^{4}$He core in $^{6}$He is ``hard''.}
We estimate the single-particle form-factor, $\langle k_2 \rangle$,  with a  wavefunction  approximated as a constant within the potential and an exponential tail outside, the decay constant of which depends  on  the $2n$ separation energy.  The coupling is seen to decrease as the orbit becomes less bound. In turn, the overall dependence of $f_{\lambda=2}$ shown in Fig.~\ref{fig:formFactor} also decreases with the tunneling parameter $\mathscr{X}$. The curves represent the two cases that can be associated with a soft- and a hard core as indicated by the $B(E2)$ strengths in W.U.
For the range of BE expected for $^{40}$Mg, $f_{\lambda=2}$ is below unity for a $B(E2)$ of 10~W.U., signaling a regime where weak particle-surface coupling dominates, in contrast to the strong coupling expected in the lighter (more bound) even-$A$ magnesium isotopes from $^{32}$Mg to $^{38}$Mg. A consequence of the weak coupling between the deformed core and the neutron halo is that the latter can be considered spherical  to all practical purposes.


In the following we apply a phenomenological Weak Coupling Model~\cite{Arima71} and the Particle-Rotor Model~\cite{ShapesAndShells} to describe the spectrum of states that can arise in such a system, and when comparing  results to  experimental data in  $^{40}$Mg we assume that the $\gamma$-ray transitions of 500(14) keV and 670(16) keV correspond to the energies of the first and second $2^+$ states, such that $2_1^+ \rightarrow 0_1^+ = 500$ keV and $2_2^+ \rightarrow 0_1^+ = 670$ keV (see Refs.~\cite{Crawford19,Crawford14}).

\subsection{ Weak Coupling Phenomenological Model}

\begin{figure}[htbp]
\centering
  \includegraphics[trim=0 60 40 60, clip,width=\columnwidth]{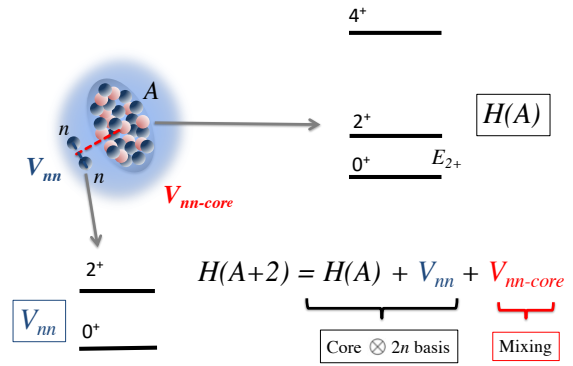}
  \caption{Sketch of our weak-coupling approach to treat the mixing of the rotational core and the two neutrons degrees of freedom, where the matrix element $V_{nn-core}$ is anticipated to be small in comparison to the 
  characteristic energies of the $2^{+}$ of  the core and the $V_{nn}$ in the $\nu p_{3/2}^2$ configuration.}
  \label{fig:couplingSketch}
\end{figure}

The  coupling of a two neutron halo (e.g.~$\nu p^2$)  to a core is schematically illustrated in Fig.~\ref{fig:couplingSketch} and the phenomenological Hamiltonian describing the relevant degrees of freedom of the total system can be written as:
\begin{equation}
\label{eq:Hamiltonian}
H(A+2)=H(A) + V_{nn} +  V_{nn-core}
\end{equation}
\noindent
The unperturbed energies of the core and two neutron subsystems are given by $H(A)$ and $V_{nn}$ respectively, and $V_{nn-core}$ is the interaction between them.

It is natural to expect that effects of weak binding on  excited states will show when the energy scales of the two degrees of freedom become comparable: 
\begin{equation}
 E_{core}(2^+) \approx  E_{2n}(2^+)
\end{equation}
As discussed in appendix A, $^{40}$Mg is one of few cases where this condition may be satisfied for $p$ neutrons outside a deformed core. 

It is interesting to note the recent work in Ref.~\cite{Ogawa21}, where the authors analyze the $2n$ decay mode of the $2^+_1$ resonant state of $^6$He.   They show that the calculated double-differential breakup cross section for the $^6$He + $^{12}$C reaction at 240 MeV/nucleon  will have a shoulder peak which comes from the  $2n$ configuration in a $\nu p^2$ $2^+$ state.

\subsubsection{State Energies and Wavefunctions}

The formation of a neutron halo is limited to weakly bound low-$\ell$ $s$ and $p$ orbits and the maximum spin for a two neutron halo system is then $2^+$.  For the lowest states in the scenario above, namely the 2$^{+}$ states, this results in a 2$\times$2 matrix that can be studied as a function of the 2n binding energy. 

\begin{figure}[htbp]
\centering
  \includegraphics[trim=40 0 40 60, clip,width=0.8\columnwidth,angle=90]{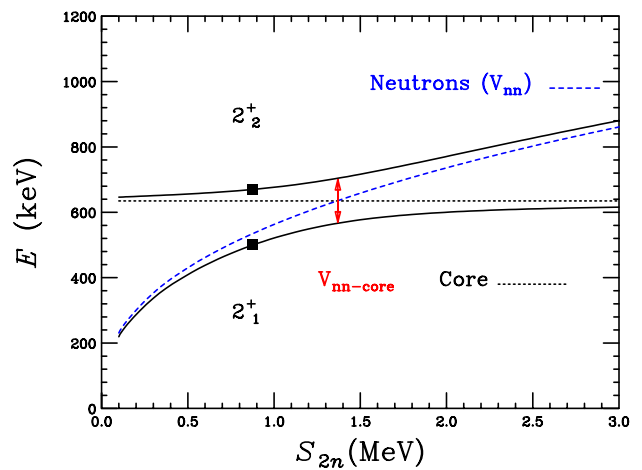}
  \caption{ Results of the weak coupling phenomenological model vs. $S_{2n}$ for the case of a $\nu p$ orbit (dashed blue line). The experimental energies are shown as black squares at the minimization solution. The two-body matrix element $V_{nn}$ is assumed to scale with volume from $^{50}$Ca.}
  \label{fig:weakCoupling}
\end{figure}

The results of the diagonalization for the 2$^{+}$ state matrix vs. 2n separation energy are shown in Fig.~\ref{fig:weakCoupling} using the example of $^{40}$Mg.  In choosing the initial parameters we  consider that the core unperturbed energy,  $E_{core}(2^+)$, is that of $^{38}$Mg and does not depend on the neutron binding (dashed black line).  The energy of the $2^+$ neutron excitation, $V_{nn}$, has been scaled by volume from that in $^{50}$Ca ($\nu p_{3/2}^{2}$ outside of $^{48}$Ca) and as expected, decreases with $S_{2n}$ (dashed blue line). It is interesting to see that the unperturbed lines cross for binding energies in the range expected for  $^{40}$Mg and even a small mixing matrix element $V_{nn-core}$ will give rise to largely mixed states in the laboratory frame. A minimization procedure on the experimental energies of the two potential states populated~\cite{Crawford19} gives a solution with $V_{nn-core}= 69$keV, $S_{2n}$=877 keV, and  wavefunctions:
\begin{eqnarray}\label{eq:WF}
|2^+_1 \rangle &=& ~0.89 |2^+_{core}\rangle + 0.45|2^+_{2n} \rangle \nonumber \\
|2^+_2 \rangle &=& -0.45 |2^+_{core}\rangle +0.89|2^+_{2n}\rangle
\end{eqnarray}
\noindent
The fact that $V_{nn-core} \ll E_{core}(2^+)~( V_{nn})$ in this solution supports the weak coupling assumption. 

\subsubsection{Reaction Cross Section}

The wavefunctions of the two $2^+$ states (see Eq.~\ref{eq:WF}) can also be used to determine their relative intensities populated in a direct knockout reaction using the procedure described in Refs.~\cite{Elbek69,Macchiavelli17,Macchiavelli18} and compared to that observed~\cite{Crawford19}. The cross-section to populate a given state in a direct reaction depends on the overlap of the initial and final state wavefunctions and may therefore be sensitive to the effects of weak binding.

\begin{figure}[htbp]
\centering
  \includegraphics[trim=0 0 0 60, clip,width=\columnwidth,angle=90]{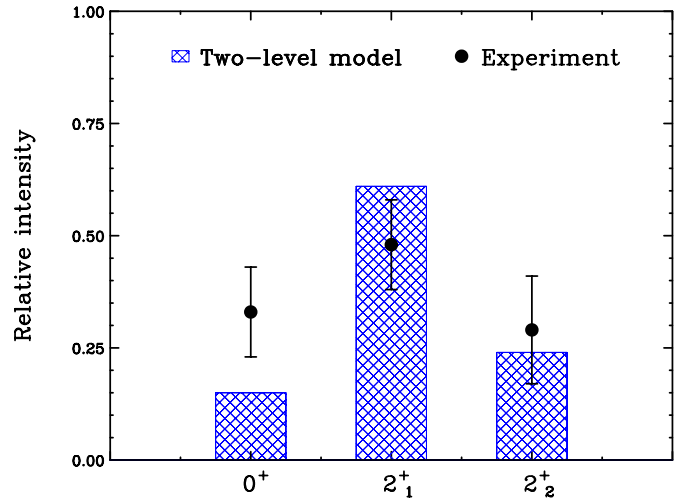}
  \caption{Minimization results for the intensity ratio of the $2^+_1$ and $2^+_2$ states populated in the experiment.}
  \label{fig:minimization}
\end{figure}

To calculate the population  of the final states in $^{40}$Mg produced from the $^{41}$Al-1p reaction we assume that the ground state of  $^{41}$Al is $K=5/2^+$, from the $\pi$[202]5/2 Nilsson level originating from the $d_{5/2}$ spherical level. In the single-$j$ approximation the collective spectroscopic factors follow the values of the Clebsch-Gordan coefficients: $\langle \frac{5}{2} \frac{5}{2} \frac{5}{2}-\frac{5}{2} |I_{f}0\rangle$. In the minimization procedure we also include a single-particle spectroscopic factor $S_{sp}(5/2^+ \rightarrow 2^+_{2n}$) with a fitted value of 0.14.
This gives a measure of the component of the $|2^+_{2n} \rangle$ state in the ground state of $^{41}$Al. A good agreement with the observed relative population of the states as determined from the $\gamma$-ray intensities~\cite{Crawford19} with the assumption of parallel decays can be seen in Fig.~\ref{fig:minimization}.

\subsubsection{Transition Probabilities}

No data currently exists on $E2$ transition probabilities ($B(E2)$s) to low-lying states in a halo nucleus and it is interesting to consider the contribution of the neutron halo to excite the $2^+_1$ and $2_2^+$ states, for example in an intermediate energy Coulomb excitation experiment~\cite{Glasmacher98}.   
\begin{figure}[htbp]
\centering
\includegraphics[trim=40 0 0 60 , clip,width=0.9\columnwidth, angle=90]{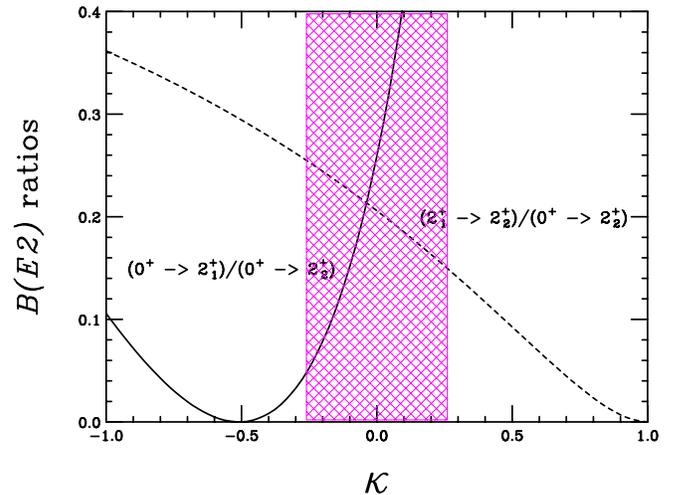}
  \caption{ Ratios of $B(E2)$s for the $0^+ \rightarrow 2^+_1$ and $2^+_1 \rightarrow 2_2^+$ transitions to the $0^+ \rightarrow 2^+_2$, as a function of the parameter $\kappa$. The (magenta) shaded region indicates the expected range for the values of $\kappa$.}
  \label{fig:BE2ratio}
\end{figure}

In the weak coupling framework we can calculate the  $B(E2)'s$ and find the ratios to excite the $2^+_1$ and $2_2^+$ states:
\begin{equation}
\frac{B(E2,0^+ \rightarrow 2^+_1)} { B(E2,0^+ \rightarrow 2^+_2)} \approx \frac{(\alpha + \beta \kappa)^2} {(-\beta + \alpha \kappa)^2}
\end{equation}
\begin{equation}
\frac{B(E2,2^+_1 \rightarrow 2^+_2)} { B(E2,0^+ \rightarrow 2^+_2)} \approx \frac{(\alpha(1 - \kappa))^2}{(1- (\alpha/\beta) \kappa)^2}
\end{equation}
\noindent
where $\kappa=\langle 0^+| \mathscr{M}(E2)_{2n} |2^+_{2n}\rangle/ \langle 0^+| \mathscr{M}(E2)_{core} |2^+_{core}\rangle  = \langle 2^+_{2n}| \mathscr{M}(E2)_{2n} |2^+_{2n}\rangle/ \langle 2^+_{core}| \mathscr{M}(E2)_{core} |2^+_{core}\rangle  $,
and $\alpha$ (=0.89) and $\beta$ (=0.45) are the amplitudes in Eq.~\ref{eq:WF}. These ratios are shown in Fig.~\ref{fig:BE2ratio}, and we note that they are sensitive to the halo ($2n$) contribution. We may anticipate that the $2n$ matrix element will be of order one single-particle unit or smaller depending on the effective charge acquire by the neutrons. In turn, the $B(E2)$ of $^{38}$Mg should be similar to that of  $^{36}$Mg~\cite{Doornenbal16}, $\approx$ 14 W.U. Thus, the shaded region indicates the expected range for $\kappa$, estimated from the values above. It is worth pointing out that from these results, the branching ratio ($2^+_2 \rightarrow 2^+_1$) for a 170 keV transition from the assumed 670 keV state is expected to be small. 

Although perhaps more experimentally ambitious, the magnetic moment of the $2_1^+$ state can also provide a sensitive measurement of the $2n$ contribution since its gyro-magnetic factor is:
\begin{equation}
g(2^+_1)=  \alpha^2 \frac{Z}{A} + \beta^2 \frac{1}{2} g_{s,\nu}
\end{equation}
giving $g(2^+_1) \approx -0.14~\mu_N$.

\subsection{Particle-Rotor Model}

The Particle-Rotor Model~\cite{ShapesAndShells} has been used extensively to describe the coupling and alignment  of valence nucleons outside a deformed core. Within this framework, the backbending phenomenon~\cite{Johnson71} was explained by Stephens~\cite{Stephens72} as a result of the crossing of the ground state ($gs$ or {\sl yrast}) band with an excited aligned band carrying  single-particle angular momenta usually referred to as the $s$ or {\sl yrare}  band.  This  can be simply understood by looking at the energies of the two bands as a function of angular momentum:
\begin{equation}
E_{gs} \approx  \frac{\hbar^{2}}{2\mathscr{I}}I^2
\end{equation}
and
\begin{equation}
E_{s} \approx  \frac{\hbar^{2}}{2\mathscr{I}}(I-2j)^2 + 2\Delta
\end{equation}
where for simplicity we assume that the two bands have equal moments of inertia, $\mathscr{I}$.  At a critical angular momentum, $I_c$, the extra energy (2$\Delta$) required to break the pair is compensated by the loss of  rotational energy due to the contribution, $2j$, of the aligned particles to the total spin.  An estimate from Eqs. 8 and 9 gives:
\begin{equation}
I_c \approx  j + \frac{3\Delta}{j^2E_{2^+}}
\end{equation}
Typically, in the rare earth, these crossings occur at relatively high-spins, while here it is anticipated to be at low-spin.

\begin{figure}[htbp]
\centering
\includegraphics[trim=80 180 100 160 , clip,width=0.9\columnwidth, angle=0]{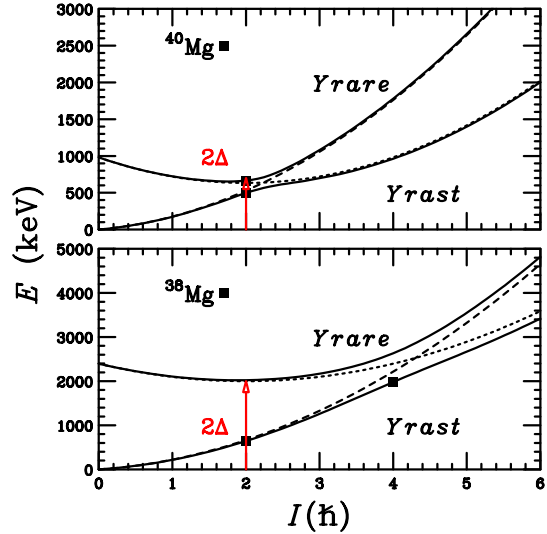}
  \caption{ Band-crossing scenario applied to both $^{38}$Mg (bottom) and $^{40}$Mg (top) showing the need for a quenched gap in $^{40}$Mg (red arrows) that can be attributed to the weak binding of the $p_{3/2}$ orbital. The $gs$ and $s$ bands are shown by the dash and dotted lines respectively}
  \label{fig:bandCrossing}
\end{figure}

In order to apply this scenario to the case of a deformed weakly bound nucleus we first look at the $^{38}$Mg {\sl yrast} band and, assuming the maximum  alignment of $2j=2$ for the $p_{3/2}$ level, we adjust the Coriolis mixing matrix element to reproduce the observed energies of the $I=2^+$ and $I=4^+$ levels in  $^{38}$Mg (Fig. \ref{fig:bandCrossing} bottom panel) from the crossing of the $gs$ (dashed line) and $s$ bands (dotted line).  

 For $^{40}$Mg, we consider the $2^+_1$ and $2^+_2$ to belong to the $gs$ ({\sl yrast}) and $s$  ({\sl yrare}) bands respectively. Assuming the same Coriolis matrix element for $^{40}$Mg as derived for $^{38}$Mg,  the similarity in their energies requires a reduced pairing gap to $2\Delta \approx 0.75$~MeV, which is half of that in $^{38}$Mg (Fig. \ref{fig:bandCrossing} top panel) to reproduce the data. That is,  the energy to break a neutron $p_{3/2}$ pair in $^{40}$Mg needs to be reduced, which is suggestive of a quenched pairing due to the reduced overlap induced by the weak binding, this quenching is consistent with the volume overlap shown in Fig.~\ref{fig:2nHalos}. It is worth  pointing out that with the parameters above,  a   {\it backbend} in angular momentum vs. rotational frequency is predicted for $^{40}$Mg. 

\section{Conclusion}
 While the phenomenon of one- and two-neutron ground-state halo nuclei is well established, the effects of weak binding on the low-lying excitation spectrum remain largely unexplored. To address this interesting question  we have studied the coupling of weakly bound (halo) valence neutrons to a deformed core using a Weak-Coupling phenomenological approach and the Particle-Rotor model. 
A “universal”plot that relates the $2n$  separation energy (S2n) to the volume overlap between the core and halo was used to characterize and identify possible halo nuclei. We illustrated our results using the known properties of $^{38,40}$Mg to discuss the impact of weak binding on the low lying excitation spectrum, one proton removal reaction cross-sections and  transition probabilities.   Despite its simplicity, our phenomenological model captures the main physical ingredients and provides a framework that allows us to to examine possible coupling schemes involving a core and halo.

Of course, other approaches to the structure of $^{40}$Mg  exist~\cite{TsunodaNature,Suzuki21} that differ in the nature of the second experimental $\gamma$ transition. It is clear that further experimental and theoretical works will be required to elucidate their intriguing structure, which we trust will be motivated by this work.

\begin{acknowledgments}

This material is based upon work supported by the U.S. Department of Energy, Office of Science, Office of Nuclear Physics under Contract No. DE-AC02-05CH11231.
AP’s work is supported in part by the Ministerio de Ciencia, Innovaci\'on y Universidades (Spain), Severo Ochoa Programme SEV-2016-0597 and grant PGC-2018-94583.
\end{acknowledgments}

\section{Appendix A}

The condition that the  energy scales of the two degrees of freedom within in weakly bound system are comparable \begin{equation}
 E_{core}(2^+) \approx  E_{2n}(2^+)
\end{equation}
\noindent
can be used to signal cases when the observed spectrum of states may by modified by the interaction of the halo and core, as discussed in the main text.
To do this we first rewrite this condition explicitly for a deformed rotor and assume the energy of the two neutron $2^{+}$ is given by a modified gap, 
\begin{equation}
 \frac{2(2+1)\hbar^2}{2\mathscr{I}} \approx   \frac{2}{3} ( 2 \widetilde{\Delta}_{2n})
\end{equation}
\noindent
where the 2/3 factor is an empirical estimate that takes into account the fact that $2^+$ two-particle states are lower than the pairing gap due to the quadrupole interaction. 
The moment of inertia is a function of the mass number, the deformation and the pairing gap~\cite{Migdal59}. We use a pairing gap (in MeV):
\begin{equation}
 \Delta \approx \frac{7.36}{A^{1/3}} ( 1 - 8.15(N-Z)^2/A^2) 
\end{equation}
\noindent
derived from the fits in Ref.~\cite{Jensen84}. From the discussions in Section II we estimate $\widetilde{\Delta}_{2n}$ by modifying the expression above by a volume correction associated to the weakly bound neutrons and obtain:

\begin{equation}
 \frac{\hbar^2}{\mathscr{I}(A,\epsilon,\Delta)} \approx 0.44\Delta (\langle r_c^2 \rangle /\langle r^2\rangle )^{3/2}
\end{equation}
\noindent
where $\langle r^2 \rangle$ depends on $S_{2n}$ through the tunneling parameter $\mathscr{X}$.  For a given deformation $\epsilon$ we can solve for the locus points in the nuclear chart, $(A,N,Z)$, that satisfy Eq. (14) and where a {\sl transition} to a weakly bound {\sl "phase"} might occur. This solution is perhaps relevant only for $p$ orbits, the lowest $(\ell,j)$ from which a $2^+$ excitation could be built and for which the low centrifugal barrier still allows the formation of a halo. The result, obtained for a notional $\epsilon = 0.4$,  limits the accessible region of interest to light masses, with that near $^{40}$Mg being the heaviest that can be reached, while heavier candidates are likely to be beyond the reach of all planned experimental facilities. 
\bibliography{weakBinding}

\end{document}